\documentclass[aps,prl,floatfix,twocolumn,tightenlines,amsmath,amssymb,nofootinbib]{revtex4-1}

\usepackage{graphicx}
\usepackage{amssymb}
\usepackage{color}
\usepackage{psfrag}
\usepackage{ifsym}
\usepackage{epstopdf}
\usepackage{dsfont}
\usepackage{amsmath}
\usepackage{multirow} 
\usepackage{amsthm}
\usepackage{dsfont}

\usepackage{soul}

\newcommand{\ket}[1] {| #1 \rangle}
\newcommand{\braket}[2] {\langle #1 | #2 \rangle}

\newcommand{\ve} {\varepsilon}

\newcommand{\one}{\leavevmode\hbox{\small1\normalsize\kern-.33em1}}

\newcommand{\ba}{\begin{eqnarray}}
\newcommand{\ea}{\end{eqnarray}}

\begin{document}
\title{Measurements on the reality of the wavefunction}
\author{M. Ringbauer$^{1,2}$, B. Duffus$^{1,2}$, C. Branciard$^{1,3}$, E. G. Cavalcanti$^{4}$, A. G. White$^{1,2}$ \& A. Fedrizzi$^{1,2}$}
\affiliation{$^1$ Centre for Engineered Quantum Systems, $^2$ Centre for Quantum Computer and Communication Technology, School of Mathematics and Physics, University of Queensland, Brisbane,   QLD 4072, Australia. \\
$^3$ Institut N\'eel, CNRS and Universit\'e Grenoble Alpes, 38042 Grenoble Cedex 9, France.\\
$^4$ School of Physics, University of Sydney, NSW 2016, Australia}

\begin{abstract}
Quantum mechanics is an outstandingly successful description of nature, underpinning fields from biology through chemistry to physics. At its heart is the quantum wavefunction, the central tool for describing quantum systems. Yet it is still unclear what the wavefunction actually is: does it merely represent our limited knowledge of a system, or is it an element of reality? Recent no-go theorems\cite{Pusey2012,Colbeck2011,hardy2013are,Patra2013no-,Aaronson2013,Colbeck2013} argued that if there was any underlying reality to start with, the wavefunction must be real. However, that conclusion relied on debatable assumptions, without which a partial knowledge interpretation can be maintained to some extent\cite{Lewis2012,Aaronson2013}. A different approach is to impose bounds on the degree to which knowledge interpretations can explain quantum phenomena, such as why we cannot perfectly distinguish non-orthogonal quantum states\cite{Barrett2014,Leifer2014,Branciard2014a}. Here we experimentally test this approach with single photons. We find that no knowledge interpretation can fully explain the indistinguishability of non-orthogonal quantum states in three and four dimensions. Assuming that some underlying reality exists, our results strengthen the view that the entire wavefunction should be real. The only alternative is to adopt more unorthodox concepts such as backwards-in-time causation, or to completely abandon any notion of objective reality.
\end{abstract}

\maketitle

\emph{``Do you really believe the moon exists only when you look at it?"} Albert Einstein's famous question encapsulates a century-old debate over the measurement problem and the very nature of the quantum wavefunction\cite{mermin1985moon}. Not all scientists---including in particular Quantum Bayesianists\cite{mermin2000QBism,caves2002quantum,Fuchs2010}---believe that our observations of the physical world can be entirely derived from an underlying objective reality. If one does however want to maintain a realist position at the quantum level, a question naturally arises: does the wavefunction directly correspond to the underlying reality, or does it only represent our partial knowledge about the real state of a quantum system?

There are compelling reasons to subscribe to the latter, \emph{epistemic} view\cite{Spekkens2007}.
If the wavefunction is a state of knowledge, seemingly inconvenient quantum phenomena such as wavefunction collapse can be explained elegantly: if the wavefunction represents knowledge, a measurement only collapses our ignorance about the real state of affairs, without necessarily changing reality; there is thus no physical collapse in the epistemic picture, but rather a reassignment of a more appropriate wavefunction in the light of new information. Another example is the indistinguishability of nonorthogonal quantum states, which in the epistemic picture can be explained by a lack of information about the actual underlying reality. It has long been an open question whether the epistemic view, together with the assumption of an underlying objective reality, is compatible with quantum measurement statistics.

Let us formalize this question. Adopting the view that some underlying objective reality exists and explains quantum predictions, we shall denote the ``real state of affairs'' that completely specifies a given physical system by $\lambda$. 
The preparation of a system in a quantum state $\ket{\psi}$ may not determine $\lambda$ uniquely; instead, it determines a classical distribution $\mu_\psi$ over the set of $\lambda$'s, describing the probability that the preparation results in a specific state $\lambda$. Quantum measurement statistics on $\ket{\psi}$ are then assumed to be recovered after averaging over $\lambda$, with the probability distribution $\mu_\psi$, which is assumed to be independent of the measurement being carried out.

A model that reproduces quantum predictions within the above framework is called an \emph{ontological model}\cite{Harrigan2010}. The states $\lambda$ are called \emph{ontic states} (or, historically, \emph{hidden variables}), while the distributions $\mu_\psi$ are called \emph{epistemic states}. Specific examples of ontological models include those that involve hidden variables in addition to a real wavefunction, as was famously suggested by Einstein-Podolsky-Rosen\cite{einstein1935cqm} to address the alleged ``incompleteness'' of quantum mechanics. Another example was formulated by John Bell in his celebrated theorem\cite{Bell1964}. Here we shall however not consider Bell's additional assumption of local causality\cite{RevModPhys.86.419}; we are rather interested in the correspondence between the wavefunction that describes the quantum state of a (single) quantum system and its possible ontic states.

If the wavefunction is itself an element of the underlying reality, then it must be specified uniquely by $\lambda$. The epistemic states (i.e. probability distributions) $\mu_\psi$ and $\mu_\phi$ corresponding to any two distinct pure states $\ket{\psi}$ and $\ket{\phi}$, Fig.~1a, must then be disjoint, Fig.~1b. An ontological model satisfying this condition is called \emph{$\psi$-ontic}.
In all other cases the wavefunction has to be treated as a representation of the limited knowledge about the real state of the system---a so-called \emph{$\psi$-epistemic} model. In such models the epistemic states of two distinct quantum states might overlap, so that a single ontic state might correspond to different pure states, Fig.~1c.

\begin{figure}[h!]
\begin{center}
\includegraphics[width=\columnwidth]{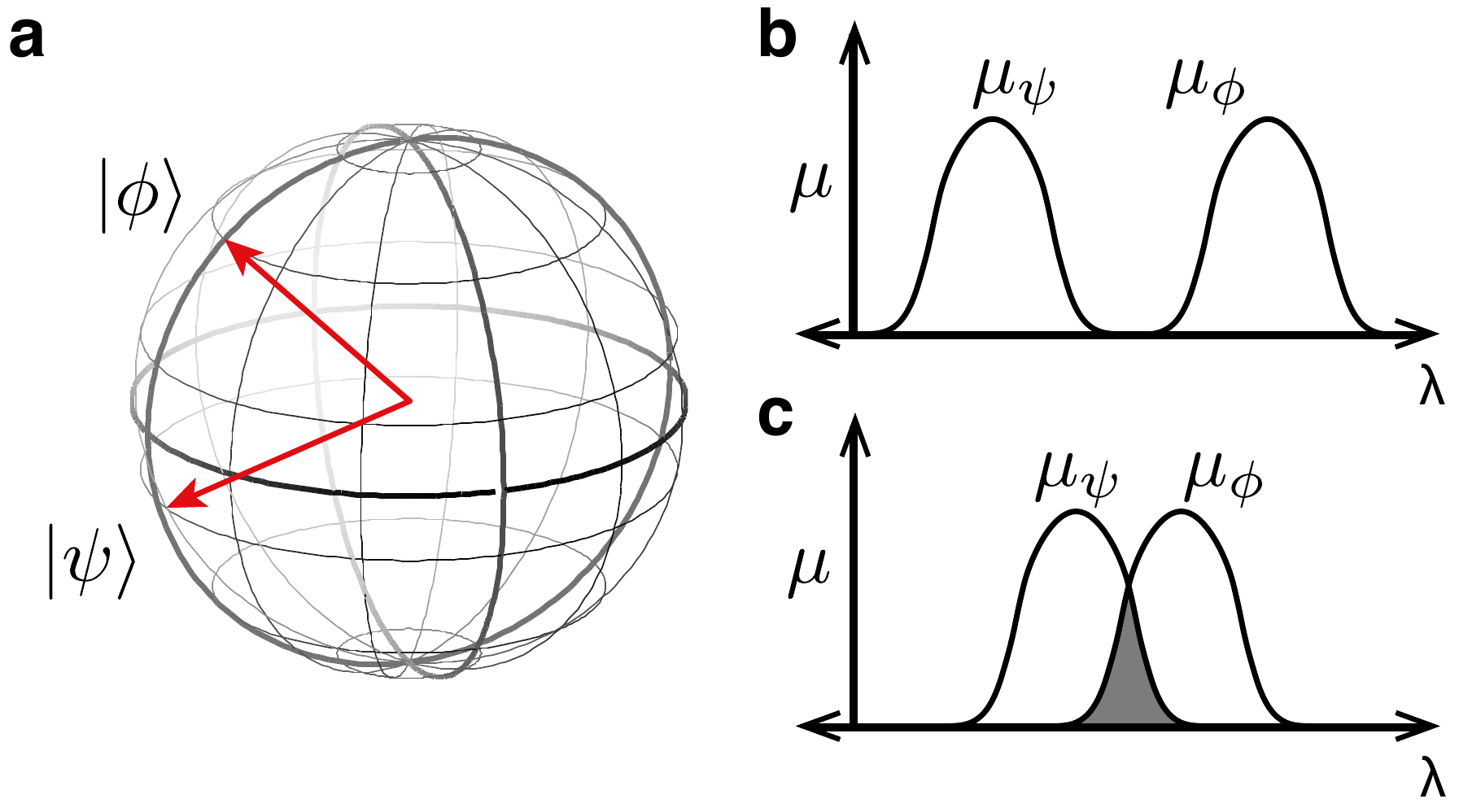}
\end{center}
\vspace{-1.5em}
\caption{\textbf{Ontological models for quantum theory.} \textbf{a} Pure quantum states $\ket{\phi}$ and $\ket{\psi}$ correspond to unit vectors in a $d$-dimensional Hilbert space.
\textbf{b-c} In ontological models every quantum state $\ket{\psi}$ is associated with a probability distribution $\mu_\psi$ over the set of ontic states $\lambda$. \textbf{b} In a $\psi$-ontic model, the distributions are disjoint for any pair of non-identical pure quantum states, such that the state itself can be regarded as an ontic element of the objective reality. \textbf{c} In $\psi$-epistemic models, the probability distributions can overlap and the quantum state is not uniquely determined by the underlying ontic state $\lambda$.}
  \label{fig:Motivation}
\end{figure}

A breakthrough in the study of these models was recently made by Pusey, Barrett and Rudolph in Ref.~\cite{Pusey2012}. They showed a no-go theorem suggesting that $\psi$-epistemic models were \emph{not} compatible with quantum mechanics. However, their theorem and related ones that followed\cite{Colbeck2011,hardy2013are,Patra2013no-,Aaronson2013,Colbeck2013} crucially relied on additional, somewhat problematic assumptions beyond the basic framework for these models.
For example, Pusey \emph{et al}. assume that independently-prepared systems have independent physical states\cite{Pusey2012}. This requirement has been challenged\cite{Emerson2013} as being analogous to Bell's local causality, which is already ruled out by Bell's theorem\cite{Bell1964}.

While completely ruling out $\psi$-epistemic models is impossible without such additional assumptions\cite{Lewis2012,Aaronson2013}, one can still severely constrain them, and bound the degree to which they can explain quantum phenomena. In particular, it was shown theoretically\cite{Barrett2014,Leifer2014,Branciard2014a} that the limited distinguishability of non-orthogonal quantum states cannot be \emph{fully} explained by $\psi$-epistemic models for systems of dimension larger than $2$.
Here we demonstrate this experimentally on indivisible quantum systems---single photons---in $3$ and $4$ dimensions. We conclusively rule out the possibility that quantum indistinguishability can be fully explained by any $\psi$-epistemic model that reproduces our observed statistics. We further establish experimental bounds on how much they can explain.
Our implementation relies on a result by Branciard\cite{Branciard2014a}, which generalises the proof of Ref.~\cite{Barrett2014}.

A $\psi$-epistemic model could explain the limited distinguishability of a pair of non-orthogonal quantum states $\ket{\psi}, \ket{\phi}$ as resulting from their two different preparation procedures sometimes producing the same ontic state $\lambda$. 
Such an explanation is fully satisfactory only if the distinguishability of two states is fully explained by the classical overlap of the probability distributions. In particular, the probability of successfully distinguishing two quantum states using optimal quantum measurements must be the same as that of distinguishing the two corresponding epistemic states, given access to the ontic states. These probabilities are given by $1-\omega_{\textsc{q}}(\ket\phi,\ket\psi)/2$ and $1-\omega_{\textsc{c}}(\mu_\phi,\mu_\psi)/2$, respectively, where $\omega_{\textsc{q}}(\ket\phi,\ket\psi)=1-\sqrt{1-|\braket{\psi}{\phi}|^2}$ is the quantum overlap of the two states and $\omega_{\textsc{c}}(\mu_\phi,\mu_\psi) = \int \min[\mu_\phi(\lambda), \mu_\psi(\lambda)] \text{d}\lambda$ is the classical overlap of the probability distributions\cite{Nigg2012,Barrett2014}. Note that $0 \leq \omega_{\textsc{c}} \leq \omega_{\textsc{q}} \leq 1$ in general\cite{Nigg2012}; a model that satisfies $\omega_{\textsc{q}}(\ket\phi,\ket\psi) = \omega_{\textsc{c}}(\mu_\phi,\mu_\psi)$ for all states $\psi,\phi$---i.e. for which the two above-mentioned success probabilities are equal and all the indistinguishability is thus explained by the overlapping probability distributions---is called \emph{maximally $\psi$-epistemic}\cite{Barrett2014}.

Maximally $\psi$-epistemic models can, however, not reproduce all quantum measurement statistics\cite{Barrett2014}.
As detailed in Ref.~\cite{Branciard2014a}, one approach to demonstrate that is to prepare a set of $n+1$ quantum states $\{\psi_j\}_{j=0}^{n}$ with $n\geq 3$, and for each triplet of states $\{\ket{\psi_0},\ket{\psi_{j_1}},\ket{\psi_{j_2}}\}$ perform a measurement $M_{j_1j_2}$ with three outcomes $(m_0,m_1,m_2)$. Denoting $j_0=0$ and $P_{M_{j_1j_2}}(m_i | \psi_{j_i})$ the probability for the outcome $m_i$, when performing the measurement $M_{j_1j_2}$ on the state $\psi_{j_i}$, the following inequality has to be satisfied by maximally $\psi$-epistemic models\cite{Branciard2014a}:
\begin{align}
S(\{\psi_j\},\{M_{j_1j_2}\}) \, = \, \frac{1+\sum\limits_{j_1<j_2} \sum\limits_{i=0}^2 P_{M_{j_1j_2}}(m_i | \psi_{j_i})}{\sum\limits_{j} \omega_{\textsc{q}}(\ket{\psi_0},\ket{\psi_j})} \ \geq \ 1 ,
\label{eq:Inequality}
\end{align}
where $1\leq j_1<j_2\leq n$ and $1\leq j\leq n$, respectively. This inequality can be violated by quantum statistics as we demonstrate below. 
Note that while Eq. (1) has a distinct Bell-inequality flavour, it aims at testing the \textit{epistemicity} of ontological models, rather than their local causality. Importantly, we are not testing the validity of quantum mechanics, but rather the compatibility of a specific interpretation with the predictions of quantum mechanics. It is therefore not surprising, but noteworthy that, in contrast to Bell inequalities, Eq. (1) is not theory-independent, in the sense that quantum theory is used explicitly to calculate the values of $\omega_{\textsc{q}}$.

For non-maximally $\psi$-epistemic models, $S$ represents an upper bound on the smallest ratio between the classical and quantum overlaps $\kappa(\psi_0,\psi_j){=}\omega_{\textsc{c}}(\mu_{\psi_0},\mu_{\psi_j})/ \omega_{\textsc{q}}(\ket{\psi_0},\ket{\psi_j})$ for all states $\psi_j \ (1\leq j\leq n)$ under consideration: the quantity $\kappa_0 = \min_j \kappa(\psi_0,\psi_j)$ then replaces the right-hand side of the inequality\cite{Branciard2014a}. This quantifies and restricts how much of the indistinguishability of non-orthogonal quantum states can be explained through overlapping classical probability distributions.

In our experiment we violate inequality~\eqref{eq:Inequality} and aim to obtain the lowest possible value for $S$. We use quantum states $d{=}3$ (qutrit) and $d{=}4$ (ququart) dimensions, prepared on single photons. These photons are created in pairs in a spontaneous parametric down-conversion process, and one photon is used as a trigger to herald the presence of a signal photon in the experiment. 

\begin{figure}[h!]
\begin{center}
\includegraphics[width=\columnwidth]{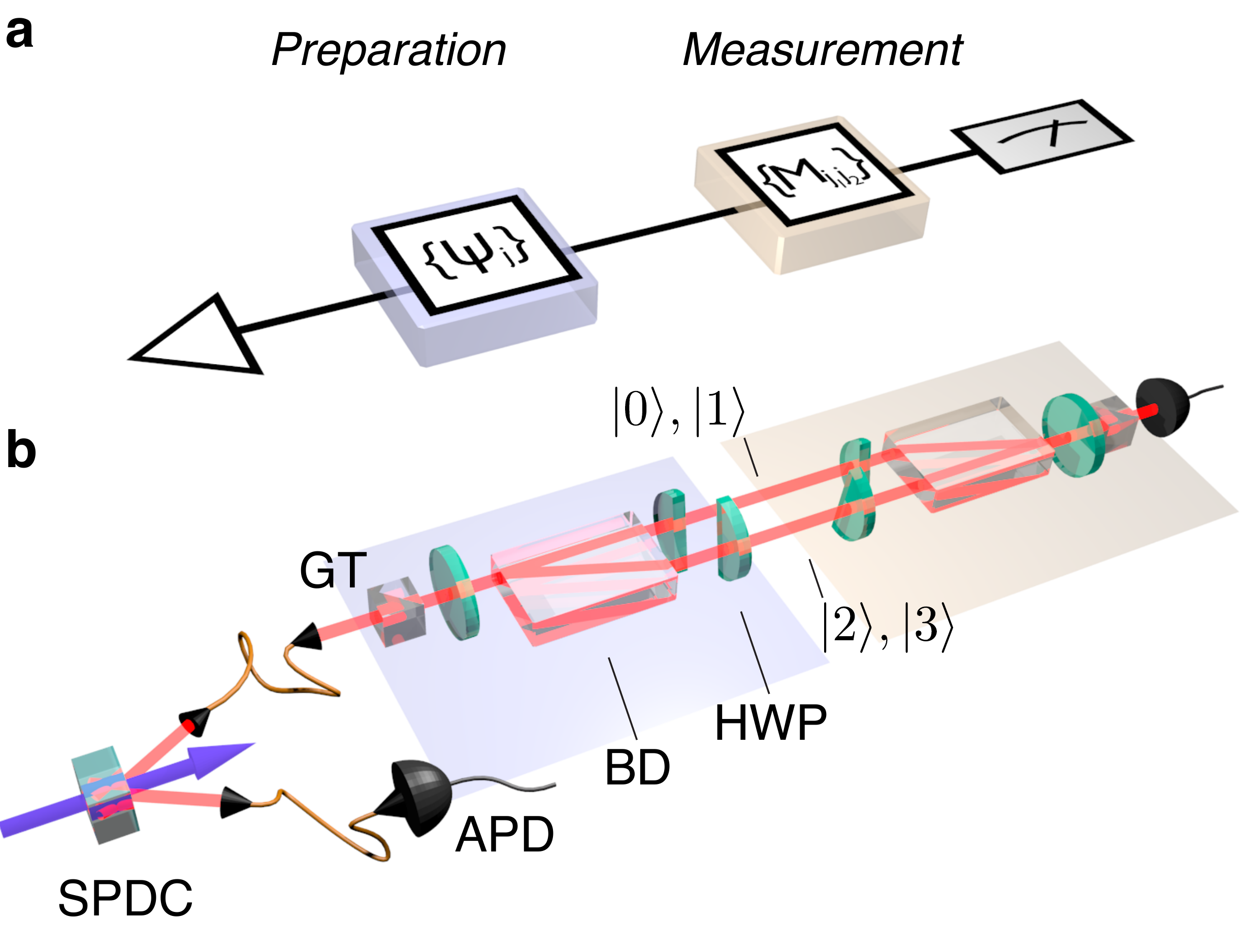}
\end{center}
\vspace{-1.5em}
\caption{\textbf{Scheme for probing the reality of the wavefunction}. \textbf{a} A $d$-dimensional system is prepared in a state from the set $\{\ket{\psi_{j}}\}$ and then subjected to measurements $\{M_{j_1j_2}\}$.
\textbf{b} Experimental implementation. Pairs of single photons are created via spontaneous parametric down-conversion (SPDC) in a periodically poled potassium titanyl phosphate (KTiOPO$_4$) crystal pumped by a $410$~nm diode laser\cite{Fedrizzi2007a}. The heralded signal photon is prepared in the initial state $\ket{H}$ by means of a Glan-Taylor polariser (GT). The subsequent half-wave plate (HWP) defines the relative amplitudes of the initially populated modes $\ket{1}$ and $\ket{2}$. A calcite beam-displacer (BD) separates the orthogonal polarisation components and a set of HWPs is used to adjust the relative amplitudes of all the basis states $\ket{0}, \ket{1}, \ket{2}$ (and $\ket 3$ for the ququart). The same setup in reverse is used to perform the measurements $\{M_{j_1j_2}\}$. Using only one output port of the final analysing polariser and one single-photon detector (APD) ensures maximal fidelity of the measurement process. Furthermore, while additional quarter-wave plates (QWPs) could be used to access the full (complex) state space, this is not necessary for the present experiment. Hence, the QWPs were not used to allow for higher accuracy.}
  \label{fig:Setup}
\end{figure}

To access higher dimensions, we dual-encode our signal photon in the polarisation and path degrees of freedom\cite{Boschi1998}, see Fig.~2. The computational basis states are $\ket{0}{=}\ket{H}_1, \ket{1}{=}\ket{V}_1, \ket{2}{=}\ket{H}_2, \ket{3}{=}\ket{V}_2$, where the index denotes the spatial mode and $H$ and $V$ correspond to horizontal and vertical polarisation, respectively. The photon polarisation was manipulated with half-wave plates and polarising beam displacers. The path degree of freedom was controlled through an interferometer formed by half-wave plates and two beam displacers. Using this setup we are able to prepare and measure arbitrary states of the form $\alpha\ket 0 + \beta \ket 1+\gamma \ket 2 +\delta \ket 3$, where $\{\alpha,\beta,\gamma,\delta\}\in\mathds{R}$. In the qutrit case the state $\ket 3$ was not populated.

While the derivation of inequality~\eqref{eq:Inequality} makes no assumptions on the measurements\cite{Branciard2014a}, it is crucial to ensure accurate preparation of the states $\ket{\psi_j}$, as these are used to calculate the quantum overlaps $\omega_{\textsc{q}}$. To this end, every optical element in the experimental setup was carefully calibrated and characterised.
For increased precision and elimination of any systematic bias in the measurement, only one output port of the setup was used and the three outcomes $m_1,m_2,m_3$ of each measurement $M_{j_1j_2}$ were obtained sequentially. This design might, however, be susceptible to imperfections in the optical components and one has to make sure that the normalised sum of the measurement operators for the three outcomes is the identity operator.
Using the calibration data for our optical elements we have bounded the average drop in measurement fidelity caused by these imperfections to $0.0007 \pm 0.0002$. The probabilities $P_{M_{j_1j_2}}(m_i | \psi_{j_i})$ were estimated from the normalised single-photon count rates. On average, we obtained $2\times 10^4$ coincidence counts per measurement, with an integration time of $10$~s per setting.

In principle\cite{Branciard2014a}, inequality~\eqref{eq:Inequality} can be violated to an arbitrary degree for $d\geq 4$, that is $S$ can range arbitrarily close to $0$, if one uses a sufficiently large number of states $n$.
However, increasing $n$ also requires a quadratic increase in the number of measurements on these states, which rapidly becomes infeasible in practice. 
Nevertheless, a violation is already possible with as few as $n{+}1{=}3{+}1$ states and is becoming more pronounced as $n$ increases. We chose to prepare up to $n{=}5$ qutrit states and $n{=}15$ ququart states, see Methods, and performed a series of $4$ independent measurements for each configuration of $d$ and $n$. 

For $n{=}5$ qutrit ($d{=}3$) states, we obtain $S{=}0.9184\pm 0.002$, violating inequality \eqref{eq:Inequality} by more than $45$ standard deviations.
For $n{=}10$ ququart ($d{=}4$) states we experimentally established $S=0.690\pm0.001$, achieving a violation by more than $250$ standard deviations.
Figure~3 shows the scaling of $S$ with the number of states and dimensions. The individual measurements agree well with the expected performance of the setup (red-shaded rectangles representing 1$\sigma$ regions) and the advantage of using higher-dimensional systems is illustrated in the inset of Fig.~3, where we compare the cases $d{=}3$ and $d{=}4$ for $n{=}3,4,5$. While there is no advantage expected for $n{=}3$ with the states we use\cite{Branciard2014a}, it is clear that $S$ decreases more quickly with $n$ for $d{=}4$.

\begin{figure}[h!]
\begin{center}
\includegraphics[width=\columnwidth]{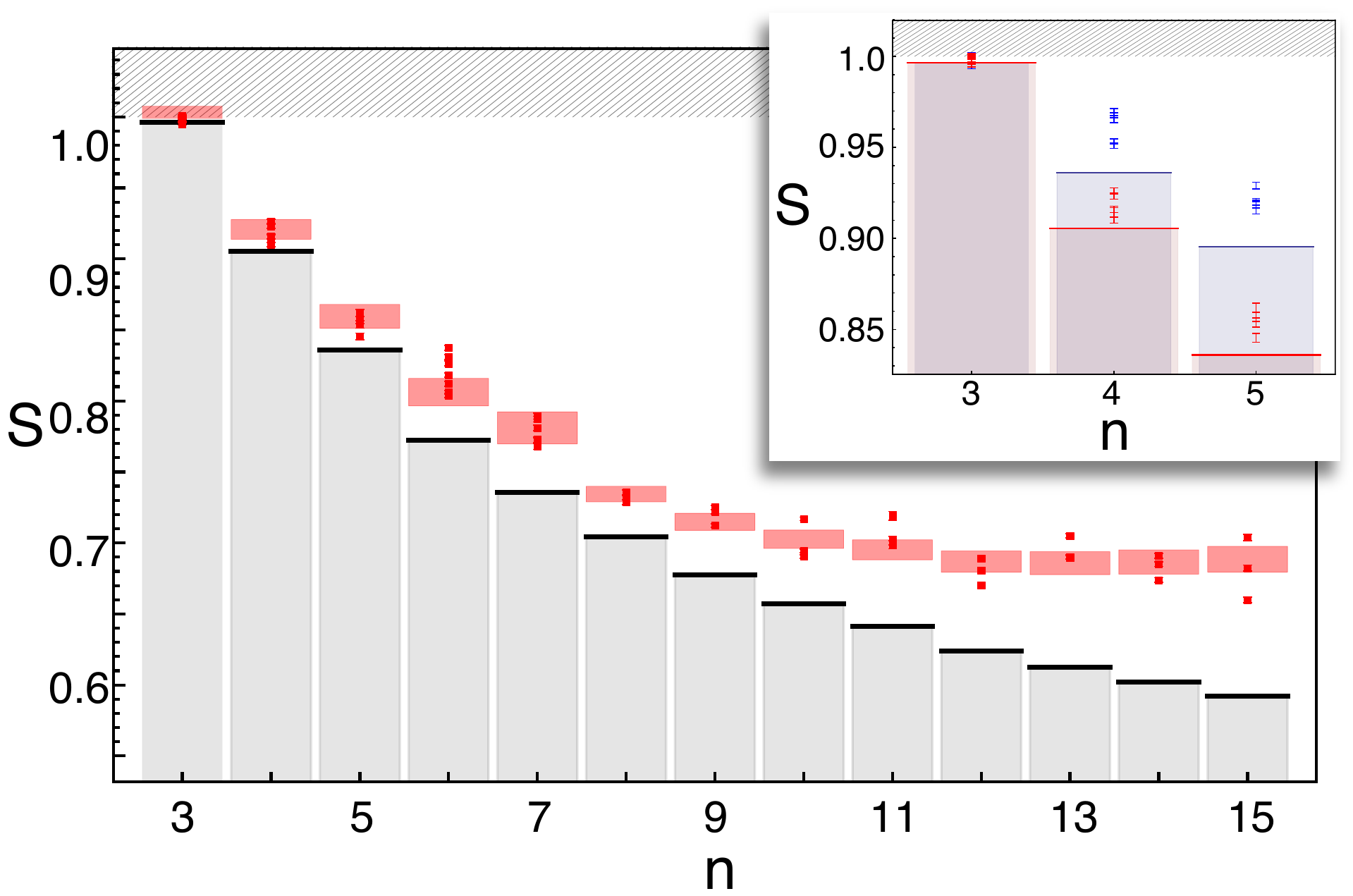}
\end{center}
\vspace{-1.5em}
\caption{\textbf{Experimental results.} \textbf{Main:} Values of $S$ for the $d{=}4$ case. \textbf{Inset:} Comparison between the $d{=}3$ (blue) and $d{=}4$ (red) cases. For both figures, the data-points represent experimental values of $S$ for an increasing number of states $n$ used in the experiment. Each data point contains a $1\sigma$ error bar, which is based on a Monte-Carlo simulation of the Poissonian counting statistics of the single-photon source, and the calibration of the various optical elements in the experimental setup, see Methods. The thick line on top of the shaded bars represent theoretical values for our states and measurements and the hatched region represents $S>1$, in which case no $\psi$-epistemic models can be ruled out. The red-shaded rectangles correspond to the $1\sigma$ range of expected values for our experimental setup based on calibration data and the observed distribution of statistical measurement errors, but do not include systematic long-term drifts of the interferometer. To illustrate this we have performed a series of at least 3 measurements for each value of $n$ and $d$. As a consequence of interferometer instability the spread between these individual runs is in general larger than the error bars and increases with the number of states $n$.}
\vspace{-1em}
  \label{fig:results}
\end{figure}
As one can see in Fig.~3, the gap between experimental and theoretical $S$-values increases with $n$, due to the compound error resulting from the quadratic increase of the number of required measurements. Eventually, a further increase of $n$ does not yield lower $S$-values. In our case the optimal trade-off is reached for $n{=}10$ and despite slightly lower experimental $S$-values no more significant violation of inequality~\eqref{eq:Inequality} could be achieved for $n>10$. This is a consequence of the extensive number of measurements and long measurement times, which cause larger fluctuations and relative errors in the data. To illustrate this we have performed at least 3 experimental runs for every data point. The spread of these values around the shaded regions increases with $n$, since the estimation of the experiment's performance does not account for long-term drifts of the interferometer. While these drifts mainly affect the numerical value of the obtained $S$, rather than its statistical significance, they might be avoided by using active stabilisation of the interferometer.
In Fig.~4 we show projections of the prepared states for the exemplary case of $n{=}7$. The deviations of the measured probabilities $P_{M_{j_1j_2}}(m_i | \psi_{j_i})$ from the expected values according to the Born rule, for one measurement in this series, ordered as they appear in Eq.~\eqref{eq:Inequality} are shown in Fig.~4b and their distribution in Fig.~4c. These data were used to calculate error bars and the expected performance of the setup. A detailed description of error handling is given in the Methods.

\begin{figure}[ht!]
\begin{center}
\includegraphics[width=\columnwidth]{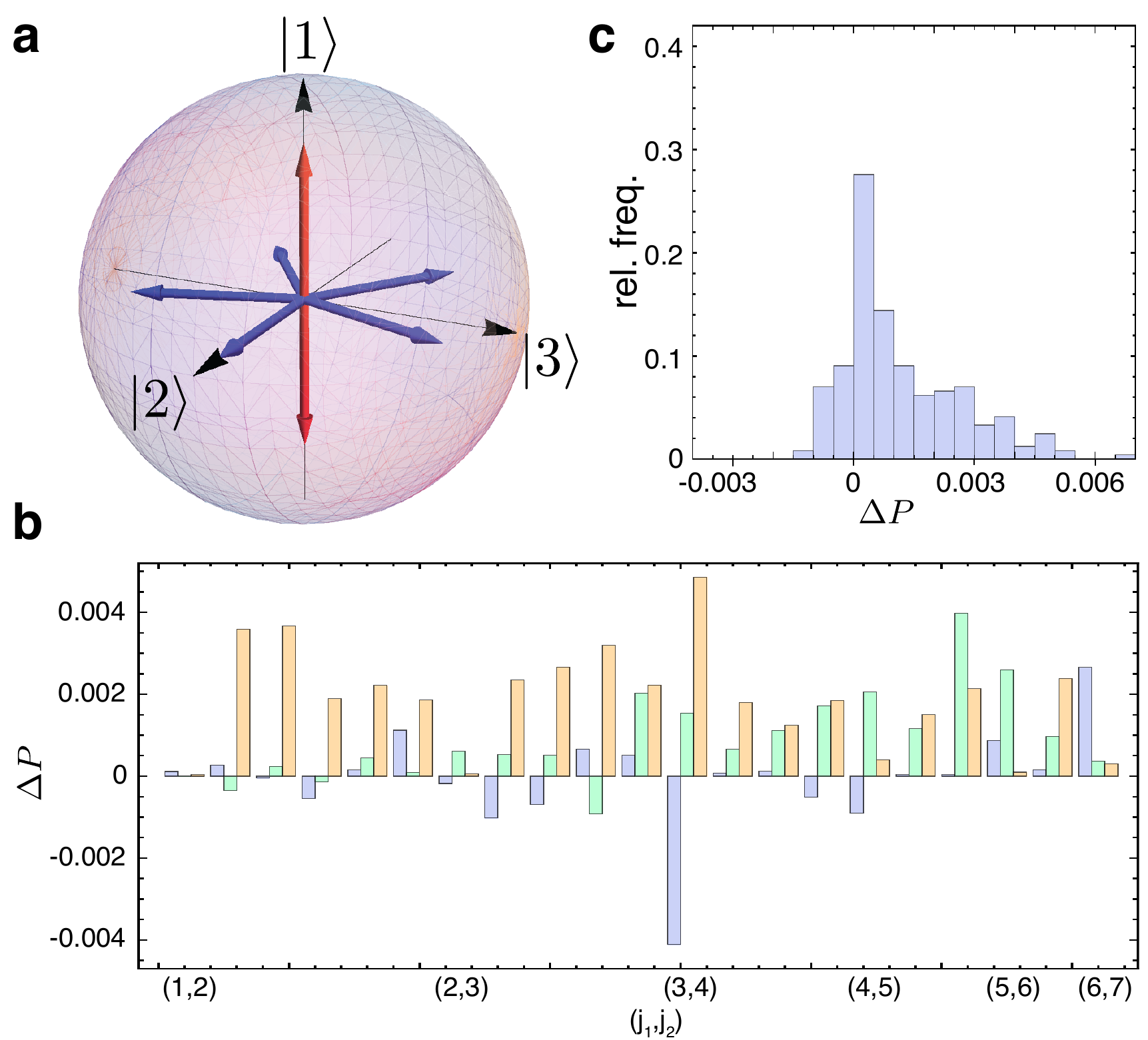}
\end{center}
\vspace{-1.5em}
\caption{\textbf{Prepared states and measurement errors for $n=7$.} \textbf{a} Projection of the ququart states $\{\ket{\psi_j}\}_{j=1}^7$ onto the three-dimensional (real) subspace orthogonal to $\ket{\psi_0}=\ket{0}$. In this example, the states are arranged quite symmetrically, pointing to the vertices of a pentagonal bi-pyramid in the latter subspace. \textbf{b} Deviations $\Delta P_{M_{j_1j_2}}(m_i | \psi_{j_i})$ of the experimentally obtained probabilities from their expected theoretical values, for the case corresponding to subfigure a. Each block of $3$ bars corresponds to one set of probabilities in Eq.~\eqref{eq:Inequality}, labeled by $(j_1,j_2)$. \textbf{c} Probability histogram of the deviations. The distribution is composed of a series of slightly shifted Gamma-distributions, which is a result of the poissonian counting statistics and comparison to theoretical expectations close or equal to $0$.}
  \label{fig:errors}
\end{figure}

Crucially, our theoretical derivation and conclusions do not require any assumptions beyond the ontological models framework---such as preparation independence\cite{Pusey2012,Nigg2012}, symmetry\cite{Aaronson2013} or continuity\cite{hardy2013are,Patra2013no-,Patra2013}---allowing us in particular to rule out a strictly larger class of $\psi$-epistemic models than the experiment of Ref.~\cite{Nigg2012}.
We do, however, rely on fair-sampling\cite{larsson2014loopholes,RevModPhys.86.419}---the physically reasonable assumption that the detected events are a fair representation of the overall ensemble---to account for optical loss and inefficient detection.
In the Methods we estimate that, with the states and measurements used in this work, an average detection efficiency above $\sim 98 \%$ would be required for ruling out maximally $\psi$-epistemic models without relying on fair-sampling. This is well above the efficiencies currently achieved in photonics, but might be achievable in other architectures such as trapped ions or superconductors, where however, precise control of qudits is yet to be demonstrated.

Recall that the ontological model framework covers all interpretations of the quantum wavefunction in which there is an observer-independent, objective reality underlying quantum mechanics. Within these realist interpretations our results conclusively rule out the most compelling $\psi$-epistemic models, namely those that fully explain quantum indistinguishability. 
They further exclude a large class of non-maximal models, characterised by a minimal ratio of classical-to-quantum overlap larger than $\kappa_0{=}0.690\pm 0.001$ 
for the states we used. Further improvements in measurement precision will allow us to impose even lower bounds on $\psi$-epistemic theories, rendering them increasingly implausible. This suggests that, if we want to hold on to objective reality, we should adopt the $\psi$-ontic viewpoint---which assigns objective reality to the wavefunction, but has some intriguing implications such as non-locality or many worlds\cite{everett1957relative}.

Alternatively, we may have to consider interpretations outside the scope of the ontological model framework, by allowing for instance retro-causality---so that the epistemic states could depend on the measurement they are subjected to---or completely abandoning any notion of observer-independent reality. This latter approach is favoured by interpretations such as QBism\cite{caves2002quantum,Fuchs2010,mermin2000QBism}, which follows Bohr's position as opposed to Einstein's.
Our work thus puts strong limitations, beyond those imposed by Bell's theorem, on possible realist interpretations of quantum theory.

\section{methods}

\subsection{Choosing states and measurements in 3 and 4 dimensions}
There is no known analytical form for the optimal states and measurements for a given $n$. We obtained $n=3, 4$ and $5$ qutrit states and $n=3,4,...,15$ ququart states numerically. To make the non-convex optimisation routine more tractable, and to allow for a precise experimental implementation, we restricted the searched state space to real vectors. These states might not be optimal; considering the full (complex) state space allows in principle for a larger violation\cite{Barrett2014,Branciard2014a} of inequality~\eqref{eq:Inequality}.
Note however that this would come with an increased experimental complexity and additional possible error sources, which may limit the advantage of using complex states.

\subsection{Error handling}
Violating inequality \eqref{eq:Inequality} requires high experimental precision, which also makes a careful error analysis essential. To obtain confidence intervals for $S$, we need to estimate the error in the quantities $P_{M_{j_1j_2}}(m_i | \psi_{j_i})$ and $\omega_{\textsc{q}}(\ket{\psi_0},\ket{\psi_j})$ in \eqref{eq:Inequality}. 

The first can be obtained directly from the measured single-photon count rates, which are Poisson distributed. Even in the most robust scenario we found, $n=5$, the maximal permissible average deviation from the predicted probabilities\cite{Branciard2014a}  $P_{M_{j_1j_2}}(m_i | \psi_{j_i})$ is $\ve_0=0.005$ for $d=3$ and $\ve_0=0.008$ for $d=4$. Our data showed an average deviation per measurement of $\ve=0.001 \pm 0.002$. In Fig.~4, we show the deviation in these measured quantities exemplary for $n=7$.

For the quantum overlaps we must account for experimentally unavoidable imperfection in the state preparation. We used independent calibration curves for our optical components to calculate a $1\sigma$ range of the systematic error in each $\omega_{\textsc{q}}(\ket{\psi_0},\ket{\psi_j})$ of the order of $10^{-3}$. We compared this estimate with results from quantum state and process tomography\cite{james2001mq}. Tomography doesn't distinguish imperfections in preparation and measurement and it likely over-estimates the actual error. We obtain an average fidelity and purity of the prepared quantum states of $\overline{\mathcal{F}}=0.998 \pm 0.002$, and $\overline{\mathcal{P}}=0.998 \pm 0.003$, respectively. We can also use this data to obtain an estimate of about $0.02$ for the standard deviation in $\omega_{\textsc{q}}$.

With an appropriate generalisation of $\omega_{\textsc{q}}$, this analysis could be extended to mixed quantum states. However, for our experiment this wasn't strictly necessary. Our downconversion source was pumped at low power to limit the probability of creating more than one photon pair within the timing resolution of the single-photon detectors to $10^{-5}$ per photon pair, and we are therefore dealing with single photons of very high intrinsic purity.

\subsection{Closing the detection loophole}
Without resorting to the fair-sampling assumption, one option to deal with no-detection events when performing a 3-outcome measurement $M_{j_1j_2}$ is to simply output a fixed or random outcome $m_i$. Suppose for simplicity that the detection efficiencies for all 3 outcomes and for all states and measurements are the same, $\eta$, and that the experimenter chooses to output $m_0$ whenever none of the detectors click. The probabilities $P_{M_{j_1j_2}}(m_0 | \psi_0)$ then become $P_{M_{j_1j_2}}^{\,(\eta)}(m_0 | \psi_0) = \eta \, P_{M_{j_1j_2}}^{(\eta=1)}(m_0 | \psi_0) + (1{-}\eta)$ and for $i \geq 1$ the probabilities $P_{M_{j_1j_2}}(m_i | \psi_{j_i})$ become $P_{M_{j_1j_2}}^{\,(\eta)}(m_i | \psi_{j_i}) = \eta \, P_{M_{j_1j_2}}^{(\eta=1)}(m_i | \psi_{j_i})$ (where $P_{M_{j_1j_2}}^{(\eta=1)}$ are the expected probabilities with perfect detection efficiency).
The value of $S$ as defined in Eq.~(1) becomes
\begin{align}
S^{\,(\eta)}(\{\psi_j\},\{M_{j_1j_2}\}) \, &= \nonumber\\
&\hspace{-6em} \frac{1+\sum\limits_{j_1<j_2} \Big[ (1{-}\eta) + \eta \sum\limits_{i=0}^2 P_{M_{j_1j_2}}^{(\eta=1)}(m_i | \psi_{j_i}) \Big]}{\sum\limits_{j} \omega_{\textsc{q}}(\ket{\psi_0},\ket{\psi_j})} 
\end{align}
(note that the same value would be obtained here if the no-detection events were replaced by random outcomes).
Ruling out maximally $\psi$-epistemic models requires one to have $S^{\,(\eta)} < 1$, i.e.
\begin{eqnarray}
\eta \, > \, \frac{1 + \frac{n(n-1)}{2} - \sum_{j} \omega_{\textsc{q}}(\ket{\psi_0},\ket{\psi_j})}{\frac{n(n-1)}{2} - \sum_{j_1<j_2} \sum_{i=0}^2 P_{M_{j_1j_2}}^{(\eta=1)}(m_i | \psi_{j_i})} .
\end{eqnarray}
For the states and measurements used in our experiment, the lowest detection efficiency threshold derived from this condition is found to be 0.976, obtained for $d=4, n=5$.

An interesting question for future research is whether the states and measurements could be optimised so as to further decrease the required detection efficiencies (recall that those used in this work were chosen to minimise the value of $S$, rather). 
Note finally that the above analysis could be refined for a practical setup to account for different detection efficiencies, by replacing for instance in each case the no-detection events by the outcome that minimises the value of $S$, or by really treating them as a fourth possible outcome.

\subsection{Acknowledgements}
We thank M.~S.~Leifer, J.~Barrett, O.~J.~E.~Maroney and R.~Lal for insightful discussions and R. Mu\~noz for experimental assistance. This work was supported in part by the Centres for Engineered Quantum Systems (Grant No. CE110001013) and for Quantum Computation and Communication Technology (Grant No. CE110001027). AGW acknowledges support from a University of Queensland Vice-Chancellor's Senior Research Fellowship, CB from the `Retour Post-Doctorants' program (ANR-13-PDOC-0026) of the French National Research Agency and a Marie Curie International Incoming Fellowship (PIIF-GA-2013-623456) of the European Commission, and CB, EGC and AF acknowledge support through Australian Research Council Discovery Early Career Researcher Awards (DE140100489, DE120100559, and DE130100240 respectively). This project was made possible through the support of a grant from Templeton World Charity Foundation, TWCF 0064/AB38.  The opinions expressed in this publication are those of the author(s) and do not necessarily reflect the views of Templeton World Charity Foundation.

\subsection{Author Contributions} 
AF, AGW, CB, EC, and MR conceived the study. AF, MR and BD designed the experiment. CB provided the lists of states and measurements to be used. MR and BD performed the experiment, collected and analysed the data. All authors contributed to writing the paper.

\end{document}